\documentclass[pdflatex,sn-nature]{sn-jnl}

\usepackage{graphicx}%
\usepackage{multirow}%
\usepackage{amsmath,amssymb,amsfonts}%
\usepackage{amsthm}%
\usepackage{mathrsfs}%
\usepackage[title]{appendix}%
\usepackage{xcolor}%
\usepackage{textcomp}%
\usepackage{manyfoot}%
\usepackage{booktabs}%
\usepackage{algorithm}%
\usepackage{algorithmicx}%
\usepackage{algpseudocode}%
\usepackage{listings}%
\theoremstyle{thmstyleone}%
\theoremstyle{thmstyletwo}%

\theoremstyle{thmstylethree}%

\begin{document}

\title[Article Title]{Reconstructing the cosmic expansion with a generalized q(z) parameterization: A decelerating Universe from late-time constraints}

\newcommand{\orcidauthorA}{0000-0003-4062-6123} 
\newcommand{\orcidauthorB}{0000-0003-0405-9344} 
\newcommand{\orcidauthorC}{0000-0002-6356-8870} 
\newcommand{\orcidauthorD}{0000-0003-1750-4769} 
\newcommand{\orcidauthorE}{0000-0003-4446-7465} 


\author*[1]{\fnm{Tom\'as} \sur{Verdugo}}\email{tomasv@astro.unam.mx}

\author[2]{\fnm{Alberto} \sur{Hern\'andez-Almada}}\email{ahalmada@uaq.mx}

\author[3]{\fnm{Miguel A.} \sur{Garc\'ia-Aspeitia}}\email{angel.garcia@ibero.mx}

\author[4]{\fnm{Juan} \sur{Maga\~na}}\email{juan.maga@ucentral.cl}

\author[5]{\fnm{Ver\'onica} \sur{Motta}}\email{veronica.motta@uv.cl}

\affil*[1]{ \orgname{Instituto de Astronom\'ia, Observatorio Astron\'omico Nacional, Universidad Nacional Aut\'onoma de M\'exico}, \orgaddress{\street{Apartado postal 106}, \city{Ensenada}, \postcode{C.P. 22800}, \country{M\'exico}}}

\affil[2]{ \orgname{Facultad de Ingenier\'ia, Universidad Aut\'onoma de
Quer\'etaro}, \orgaddress{\street{Centro Universitario Cerro de las Campanas}, \city{Santiago de 
Quer\'etaro}, \postcode{76010}, \state{Queretaro}, \country{M\'exico}}}

\affil[3]{\orgdiv{Depto. de F\'isica y Matem\'aticas}, \orgname{Universidad Iberoamericana Ciudad de M\'exico}, \orgaddress{\street{Prolongaci\'on Paseo de la Reforma 880}, \city{M\'exico}, \postcode{01219}, \state{State}, \country{M\'exico}}}

\affil[4]{\orgdiv{Escuela de Ingenier\'ia}, \orgname{Universidad Central de Chile}, \orgaddress{\street{Avenida Francisco de Aguirre 0405}, \city{ La Serena}, \postcode{171-0164}, \state{Coquimbo}, \country{Chile}}}

\affil[3]{\orgdiv{Instituto de F\'isica y Astronom\'ia}, \orgname{Universidad de Valpara\'iso}, \orgaddress{\street{Avda. Gran Breta\~na 1111}, \city{Valpara\'iso}, \postcode{2360102},  \country{Chile}}}


\abstract{We present a generalized phenomenological parameterization of the deceleration parameter $q(z)$ that incorporates an effective radiative component (ERC) in addition to a localized late-time contribution. The proposed framework extends previous two-parameter $q(z)$ reconstructions by explicitly regulating the high-redshift behavior while preserving the late-time transition dynamics. We constrain the free parameters $(h, q_0, z_c, z_e)$ using late-time observational data from cosmic chronometers (CC), Pantheon+ Type Ia supernovae (SNIa), H\,\textsc{ii} galaxies (HIIG), and intermediate-luminosity quasars (QSO). For the full data combination (CC+SNIa+HIIG+QSO), we obtain $q_0 = -0.25^{+0.04}_{-0.04}$ and a transition redshift $z_T \simeq 0.80$, indicating a currently accelerating Universe with a transition occurring earlier than in the $\Lambda$CDM model. Within the redshift range probed by the data, the reconstructed $q(z)$ deviates from the $\Lambda$CDM trend, suggesting a possible reduction of the late-time acceleration. Furthermore, the reconstruction favors a relatively high value of the Hubble parameter, $h = 0.729 \pm 0.006$. The ERC remains weakly constrained by late-time data but ensures a smooth and monotonic evolution of $q(z)$, $j(z)$, and $w_{\rm eff}(z)$ across a wide redshift range. Within the observed interval, the model effectively reproduces the late-time behavior of the previous parametrization, while providing a controlled extension toward early epochs. Our results show that current low- and intermediate-redshift data are compatible with a reduced late-time acceleration.}





\maketitle

\section{Introduction}
\label{intro}

Modern cosmology is currently undergoing a particularly dynamic phase, as contemporary observations increasingly indicate potential tensions and inconsistencies within the standard cosmological paradigm, the $\Lambda$-Cold Dark Matter model ($\Lambda$CDM) \cite{2025PDU....4901965D}. Within this framework, the dominant contribution to the total energy budget of the universe arises from dark matter, which governs the formation and evolution of large-scale structure, and dark energy, modeled as a cosmological constant, which accounts for the observed late-time accelerated expansion of the cosmos.

Despite the remarkable empirical success of the $\Lambda$CDM paradigm, several persistent tensions remain. A prominent example concerns the value of the cosmological constant: within quantum field theory, naive estimates of the vacuum energy density exceed the observed value by approximately 120 orders of magnitude \cite{Weinberg,Zeldovich}. Another significant challenge is the Hubble tension, i.e. the discrepancy between late-time (low-redshift) and early-time (high-redshift) determinations of $H_0$ \cite{DiValentino:2020naf}, which continues to resist a satisfactory theoretical explanation. More recently, observations from the Dark Energy Spectroscopic Instrument (DESI) \cite{DESI:2024hhd} suggest an expansion history of the Universe that deviates from the $\Lambda$CDM prediction, favoring a dynamical dark energy component instead of a strictly constant cosmological term. In particular, the DESI results indicate that the effective dark energy equation of state may evolve with redshift, transitioning between phantom, cosmological constant, and quintessence-like regimes as $z$ increases \cite{DESI:2025zgx}.

These cumulative tensions have motivated the community to investigate alternative frameworks capable of ameliorating or resolving the cracks in the standard cosmological model. Emergent dark energy scenarios, for instance, provide competitive explanations and, in many cases, involve no more free parameters than $\Lambda$CDM \cite{PEDE:2019ApJ,Esteban-Gutierrez:2024rpz}. Other proposals predict potentially observable imprints of past dark energy phases \cite{Alcantara-Perez:2023jbv}, remaining compatible with dynamical dark energy models. Likewise, modified gravity theories such as $f(Q)$ and $f(R)$ represent viable extensions of General Relativity that naturally accommodate a late-time accelerated expansion of the Universe \cite{Sotiriou:2008rp,Heisenberg:2023lru}. Additional problems, as well as generic proposals for models beyond $\Lambda$CDM, are reviewed in \cite{Motta:2021hvl,CosmoVerseNetwork:2025alb}.

On the other hand, an alternative framework to study cosmic dynamics is through the deceleration parameter, defined as $q \equiv -\ddot a / aH^2$, which characterizes an accelerating Universe when $q<0$ \cite{Turner:2001mx, Bolotin:2015dja} Therefore, studying its present value, $q(z=0)=q_0$, is crucial for shedding light on the nature of dark energy and the ultimate fate of the Universe. In the standard cosmological model, $q_0 \approx -1$, indicating a phase currently accelerating. However, some studies within the context of dynamical dark energy suggest that the present expansion could be non-accelerating. In particular, some analyzes have proposed that Type Ia supernova (SNIa) luminosities may depend on the properties of their progenitor systems, with important implications for their use as standard candles \cite[e.g.,][]{2013A&A...560A..66R, 2020A&A...644A.176R}. Under this hypothesis, it has been argued that the present-day deceleration parameter could satisfy $q_0 \geqslant 0$, implying a non-accelerating Universe at the current epoch \cite{Son:2025rdz}. A combined analysis of the BAO, CMB, and DES5Y data yields $q_0 = 0.178 \pm 0.061$ \cite{Son:2025rdz}, thus contributing to the current discussion on the nature of cosmic acceleration. 

These results further motivate the exploration of parametric approaches to $q(z)$ as a model-independent framework to investigate the expansion history of the Universe and to constrain the properties of the dark energy component. Several parametrizations have been proposed in the literature, including linear, quadratic, logarithmic, and divergence-free forms \cite{Gong:2006gs,Mamon:2016dlv,Bolotin:2015dja,Bolotin:2015bba,delCampo:2012ya}. In particular, kinematic reconstructions based on the deceleration parameter have been widely used to characterize the transition from decelerated to accelerated expansion without assuming a specific underlying gravitational theory \cite{2008MNRAS.390..210C,2009PhRvD..79d7301C,2008MPLA...23.1939X,2011APh....35...17S,2012JCAP...01..018N}. These methods rely solely on the homogeneous and isotropic FLRW geometry and on the definition of kinematic quantities derived from the scale factor, allowing the expansion history to be inferred directly from observational data independently of the dynamical equations governing gravity. They provide a direct way to reconstruct the expansion history using observational probes such as Type Ia supernovae, cosmic chronometers, and baryon acoustic oscillations. Similar parametrizations have also been employed within modified gravity frameworks, where the functional form of the deceleration parameter or the expansion rate is either derived or adopted in theories such as $f(R,L_m)$ \cite{10.1016/j.dark.2024.101545}, $f(Q,T)$ gravity \cite{10.1142/S0219887824500543}, and teleparallel formalisms including $f(T)$ \cite{10.1142/S0219887824501664} and $f(T, T_G)$ gravity \cite{10.1142/S0219887826501409}. In these approaches, the deceleration parameter serves as a useful diagnostic to characterize the cosmic expansion history and its transition from deceleration to acceleration within different theoretical settings. Collectively, these works highlight both the flexibility of parametric descriptions and the importance of carefully assessing their behavior, particularly at high redshift, where the reconstruction may become sensitive to the assumed functional form.

\cite{RomanGarza2019} introduced a two-parameter form of $q(z)$ capable of describing the late-time transition to cosmic acceleration, allowing for both negative and positive present-day values of the deceleration parameter, $q_0$, while neglecting the radiation component dominant at early times. Using  SNIa data \cite{2014A&A...568A..22B, 2018ApJ...859..101S} and cosmic chronometer data \cite{2018MNRAS.476.1036M}, the authors constrained the parameters of the model and obtained a value for $q_0$ consistent with the standard cosmological model. Furthermore, the reconstructed equation of state (EoS) was found to cross the phantom divide, pointing toward quintom-like dark energy behavior. In this work, we revisit and generalize that framework for the $q$ parameter by incorporating the different dynamical behaviors that characterize distinct cosmic epochs, namely radiation domination at early times, matter domination at intermediate redshifts, and dark energy domination at late times. Specifically, the main difference with respect to previous works lies in the introduction of an Effective Radiative Component (ERC), a purely kinematic contribution designed to regulate the early-time behavior of the deceleration parameter.

The outline of the paper is as follows:  the first part \ref{sec:framework} is devoted to the study of the theoretical framework, centering our attention on the deceleration parameter and deducing the respective dimensionless Friedmann equation, and additionally, we present the effective equation of state of the model. Section \ref{sec:ObsData} presents the observational data that we use to constrain our model. Section \ref{sec:Results} is dedicated to discussing all the results obtained from the constraints, and finally, in Sec. \ref{sec:Conclusions} we give our conclusions and discussion. All our equations are expressed in natural units where $c=\hbar=k_B=1$, unless we say otherwise.

\section{Theoretical framework}
\label{sec:framework}

\subsection{Functional form of the deceleration parameter}

The deceleration parameter provides a purely kinematic and model-independent
characterization of the cosmic expansion history. In this work, we adopt a
phenomenological parameterization that allows for a continuous description of 
late-time acceleration and asymptotic $q(z)$ behavior at high redshift. The parametrization adopted in this work follows a phenomenological approach similar to that commonly used in kinematic reconstructions of the expansion history  \cite[e.g.,][]{2008MNRAS.390..210C,2009PhRvD..79d7301C,2008MPLA...23.1939X,2011APh....35...17S,2012JCAP...01..018N}. Its functional form is designed to be well behaved over the redshift range of interest, allowing for a smooth transition between decelerated and accelerated phases, as well as sufficient flexibility to accommodate different possible evolutionary behaviors of the deceleration parameter. An important advantage of this choice is that it facilitates the analytical reconstruction of cosmological quantities such as $H(z)$ and derived parameters, while keeping the number of free parameters minimal. In addition, the parametrization allows for variations in the concavity of $q(z)$ and does not impose a priori restrictions on the present-day acceleration state. However, as in other parametric approaches, the behavior at high redshift is mainly driven by the assumed functional form rather than directly constrained by the data (see Section \ref{sec:ObsData}). For this reason, the physical interpretation of the results is restricted to the redshift range where observational data provide meaningful constraints, and caution must be exercised when extrapolating beyond this regime. We propose a decomposition of the deceleration parameter
into three contributions,

\begin{equation}
q(z) = \frac{1}{2} + f_{\mathrm{DE}}(z) + g_{\mathrm{ERC}}(z),
\label{eq:q_split}
\end{equation}
where the constant term $1/2$ corresponds to the matter-dominated expansion regime, $f_{\mathrm{DE}}(z)$ describes a localized late-time transition associated with an effective dark energy contribution, $g_{\mathrm{ERC}}(z)$ is an ERC,
introduced to regulate the early-time behavior of the expansion.

 Although the ERC is explicitly included in the parametrization, its impact on late-time expansion is expected to be subdominant for low-redshift data, so that the model effectively reduces to the original parametrization by \cite{RomanGarza2019} within the observed range.

Regarding both terms in Eq. \eqref{eq:q_split}, we start with the late-time contribution modeled as a localized transition with a Gaussian
profile modulated by a factor $(1+z)$, having the following equation
\begin{equation}
f_{\mathrm{DE}}(z) =
\left(q_0 - \frac{1}{2}\right)(1+z)\exp\!\left[\beta\left(z_c^2-(z+z_c)^2\right)\right],
\label{eq:fde}
\end{equation}
where $q_0$ is the current value of the deceleration parameter. The parameter $z_c$ sets the characteristic redshift around which the late-time transition governed by $f_{\rm DE}(z)$ is centered, while the actual transition redshift $z_T$ is defined implicitly by the condition $q(z_T)=0$. The parameter $\beta$ controls the width of the late-time transition encoded in the dark-energy contribution $f_{\mathrm{DE}}(z)$.  Higher values of $\beta$ correspond to a sharper and more localized transition, while smaller values produce a broader and smoother evolution in the redshift. Therefore, $\beta$ regulates how rapidly the deceleration parameter departs from the regime where matter dominates and approaches its late-time accelerating behavior.

From a kinematic perspective, $\beta$ does not introduce a new dynamical degree
of freedom, but rather parametrizes the effective rate at which the transition
occurs around the characteristic redshift $z_c$. We adopt
a tied configuration in which $\beta$ is linked to the early-time scale $z_e$ (see below)
according to $\beta = [\log_{10} z_e]^{-1}$. Note that if the ERC contribution is turned off, i.e. $g_{\rm ERC}(z)\equiv 0$,  the
transition-width parameter is fixed to $\beta=1$ by construction. This choice reproduces the original phenomenological parametrization proposed by
Rom\'an--Garza et al.~\cite{RomanGarza2019}, which describes the late-time transition
using a fixed-width profile and does not include an explicit early-time term.

Additionally, the effective radiative component is defined as
\begin{equation}
g_{\mathrm{ERC}}(z) =
\frac{1}{2}\,\frac{z}{1+z+z_e},
\label{eq:g_erc}
\end{equation}
where $z_e$ sets the redshift scale at which the behavior in the early-time becomes
relevant. This term is purely kinematic in nature. Its role is to enforce the correct asymptotic limit: $q(z) \longrightarrow 1$ $(z \gg z_e)$, corresponding to a radiation expansion regime at high redshift, while leaving the late-time dynamics essentially unaffected.

Combining Eqs.~\eqref{eq:q_split}, \eqref{eq:fde} and \eqref{eq:g_erc}, the final
expression for the deceleration parameter reads

\begin{eqnarray}
&&q(z) = \frac{1}{2}+ \left(q_0 - \frac{1}{2}\right)(1+z)
\nonumber\\
&&\times
\exp\!\left[\beta\left(z_c^2-(z+z_c)^2\right)\right]
+ \frac{1}{2}\,\frac{z}{1+z+z_e}.
\label{eq:q_final}
\end{eqnarray}
Equation~\eqref{eq:q_final} constitutes the fundamental ansatz of the
model and defines the quantity that is being parametrized and constrained. The construction adopted here is phenomenological and gives a flexible and well behaved parametrization that captures both early- and late-time features of the cosmic expansion
history. This approach yields analytically tractable expressions and stable and
numerical reconstructions. A complementary kinematic interpretation of the functional form adopted, based on the relation between deceleration and jerk parameters, is presented in Appendix~\ref{app:motivation}.

\subsection{Dimensionless Friedmann equation}

Given the deceleration parameter, the dimensionless Hubble function
$E(z)\equiv H(z)/H_0$ follows from the exact kinematic relation
\begin{equation}
\ln E(z) =
\int_0^{z}
\frac{1+q(z')}{1+z'}\,dz'.
\label{eq:lnE}
\end{equation}
For the parametrization given in Eq.~\eqref{eq:q_final}, the integral in
Eq.~\eqref{eq:lnE} can be performed analytically.
This yields a closed-form expression for the normalized Hubble rate 
$E(z)$, which is used throughout the analysis to compute
the expansion history and all derived observables. Thus, integrating Eq. \eqref{eq:lnE} with \eqref{eq:q_final} we have

\begin{eqnarray}
\ln E(z)
&=& \frac{3}{2}\ln(1+z)
+ \xi_\beta \Bigl[
\mathrm{erf}\!\left(\sqrt{\beta}(z+z_c)\right)
\nonumber\\
&&\hspace{1.6em}
- \mathrm{erf}\!\left(\sqrt{\beta}z_c\right)
\Bigr]
- \frac{1}{2 z_e}\,\ln(1+z)
\nonumber\\
&& + \frac{z_e+1}{2 z_e}
\ln\!\left(\frac{1+z+z_e}{1+z_e}\right),
\label{eq:H_final}
\end{eqnarray}

where the coefficient $\xi_\beta$ is defined as
\begin{equation}
\xi_\beta =
\frac12\sqrt{\frac{\pi}{\beta}}
\left(q_0-\frac12\right)
\exp\!\left(\beta z_c^2\right).
\end{equation}
and $\beta = [\log_{10} z_e]^{-1}$ as mentioned previously.

Although the model is defined through $q(z)$, it is important to stress
that the quantities directly compared with the observational data are the Hubble
function $H(z)$ and the corresponding distance measurements derived from it.

In addition, it is possible to reconstruct an effective
equation of state (EoS) parameter $\omega_{\rm eff}(z)$ in the form

\begin{equation}
w_{\rm eff}(z) =\frac{2}{3}[f_{\mathrm{DE}}(z) + g_{\mathrm{ERC}}(z)], \label{eq:omega_final}
\end{equation}
where Eq.~\eqref{eq:omega_final} provides a fully  reconstruction of the dark energy equation of state when radiation is explicitly included.

Finally, we also computed the jerk parameter as:
\begin{equation}
\label{eq:j(z)}
\begin{split}
j(z) ={} & [1 + 2f_{\mathrm{DE}}(z) + 2g_{\mathrm{ERC}}(z)](1 + f_{\mathrm{DE}}(z) + g_{\mathrm{ERC}}(z)) \\
& + (1+z)\frac{d}{dz}(f_{\mathrm{DE}}(z) + g_{\mathrm{ERC}}(z)),
\end{split}
\end{equation}
which indicates whether we are dealing with variable or constant dark energy.

Within this parametric framework, two characteristic redshifts naturally emerge in the cosmic evolution: the traditional transition redshift ($z_T$) at which the Universe evolves from a decelerating to an accelerating phase, and another redshift scale associated with the epoch at which the radiation component becomes dynamically relevant ($z_e$). As a result, the effective equation of state governing the cosmic expansion consistently accounts for these effects and can be characterized in terms of the values associated with the underlying kinematic components.

\begin{table*}[ht!]
\centering
\caption{Median values for the cosmological parameters and their $1\sigma$ confidence interval.}
\label{tab:bf_model}
\resizebox{0.98\textwidth}{!}{%
	\begin{tabular}{lccccccc} 
    \hline
    Data & $\chi^2$ & $h$ & $q_0$ & $z_e$  & $z_c $ & $\tau_U \,[\rm{Gyrs}]$  & $z_T $   \\ [0.9ex] 
    \hline
    \multicolumn{8}{c}{} \\ [0.9ex]
CC+SNIa          &2003.39 & $0.675^{+0.025}_{-0.025}$  & $-0.210^{+0.044}_{-0.052}$  & $14.024^{+4.152}_{-5.043}$  & $0.223^{+0.180}_{-0.137}$  & $13.073^{+0.445}_{-0.417}$  & $0.776^{+0.113}_{-0.106}$  \\ [0.9ex] 
CC+SNIa+HIIG     &2441.57 & $0.692^{+0.013}_{-0.013}$  & $-0.203^{+0.039}_{-0.047}$  & $14.830^{+3.586}_{-4.546}$  & $0.159^{+0.145}_{-0.102}$  & $12.967^{+0.271}_{-0.265}$  & $0.859^{+0.099}_{-0.099}$  \\ [0.9ex] 
CC+SNIa+HIIG+QSO &5571.86 & $0.729^{+0.006}_{-0.006}$  & $-0.250^{+0.035}_{-0.038}$  & $8.130^{+5.171}_{-2.256}$   & $0.103^{+0.148}_{-0.071}$  & $12.258^{+0.142}_{-0.137}$  & $0.801^{+0.054}_{-0.049}$  \\ [0.9ex] 
\hline
\end{tabular}%
}
\end{table*}

\begin{figure*}
    \centering
    \includegraphics[width=0.8\textwidth]{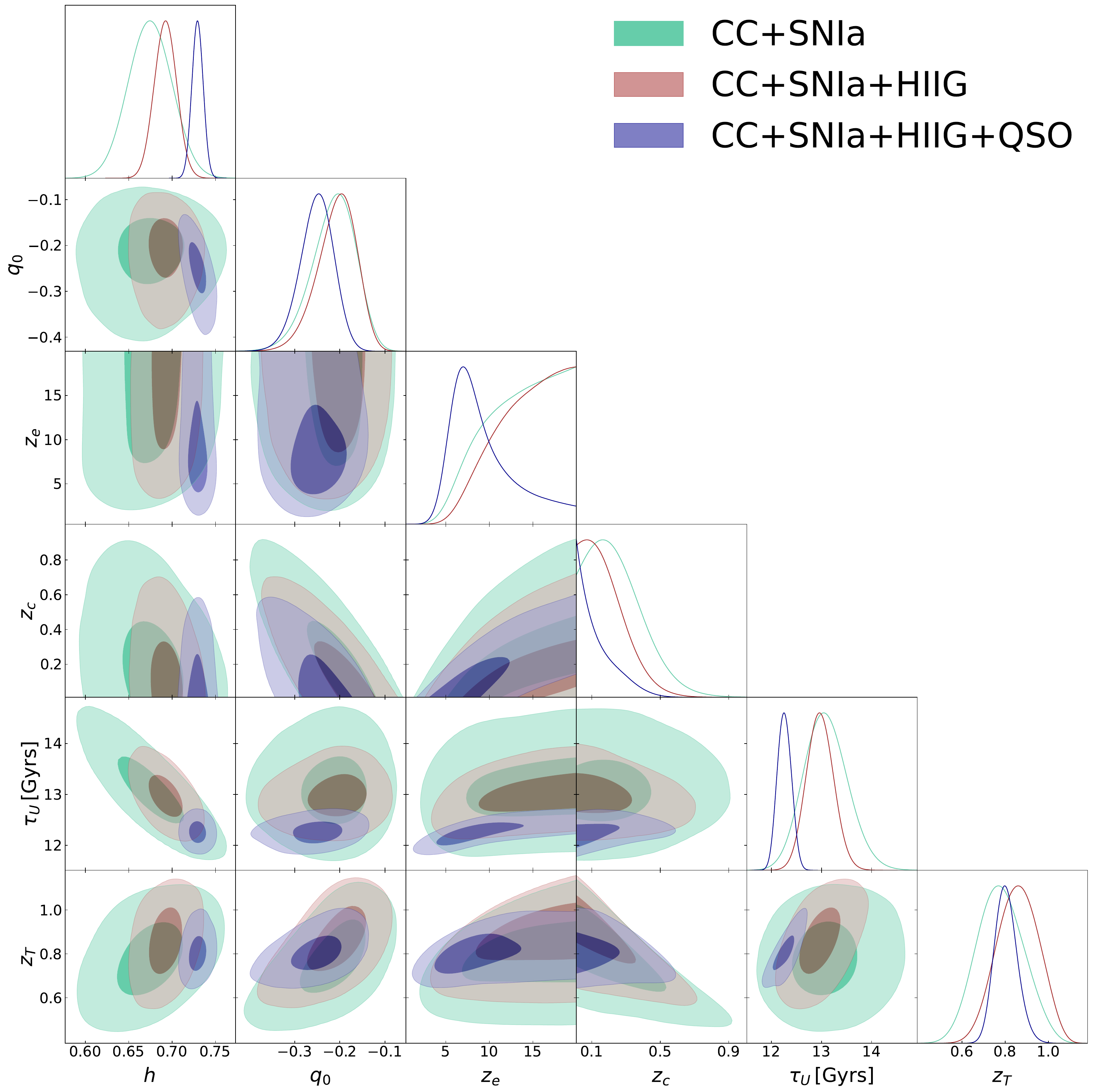}
    \caption{Posterior distributions for the kinematic $q(z)$ model parameters $\boldsymbol{\Theta}=(h,q_0,z_e,z_c)$ obtained from the MCMC analysis. Diagonal panels show the marginalized 1D posteriors, while off-diagonal panels display the joint 2D regions at $1\sigma$  and $3\sigma$ confidence levels.}
    \label{fig:contours}
\end{figure*}

\begin{figure*}
    \centering
    \includegraphics[width=0.31\textwidth]{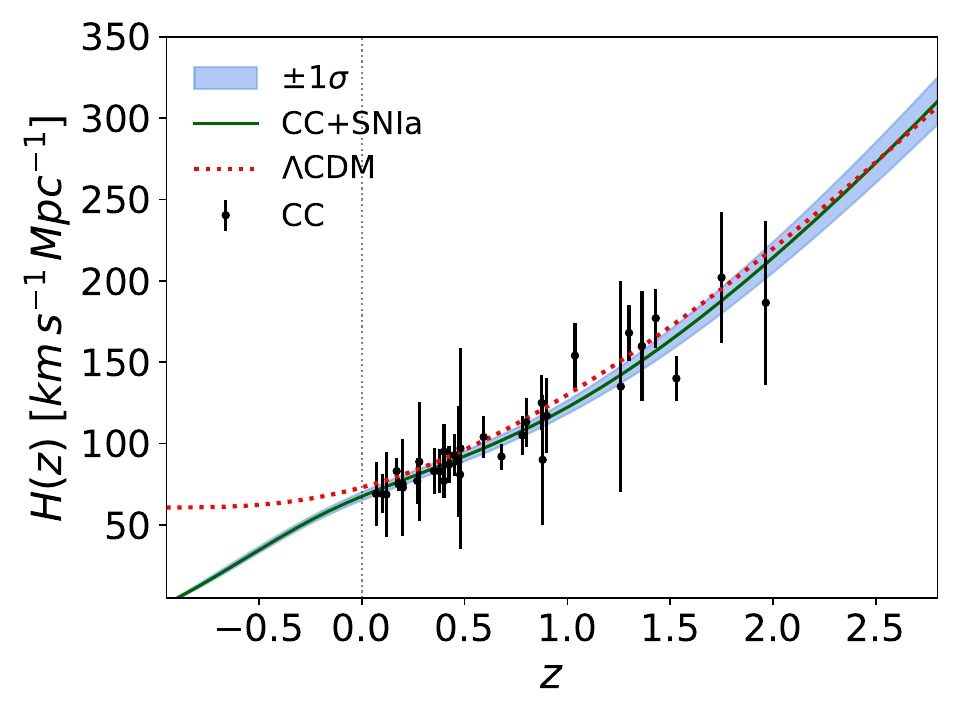}
    \includegraphics[width=0.31\textwidth]{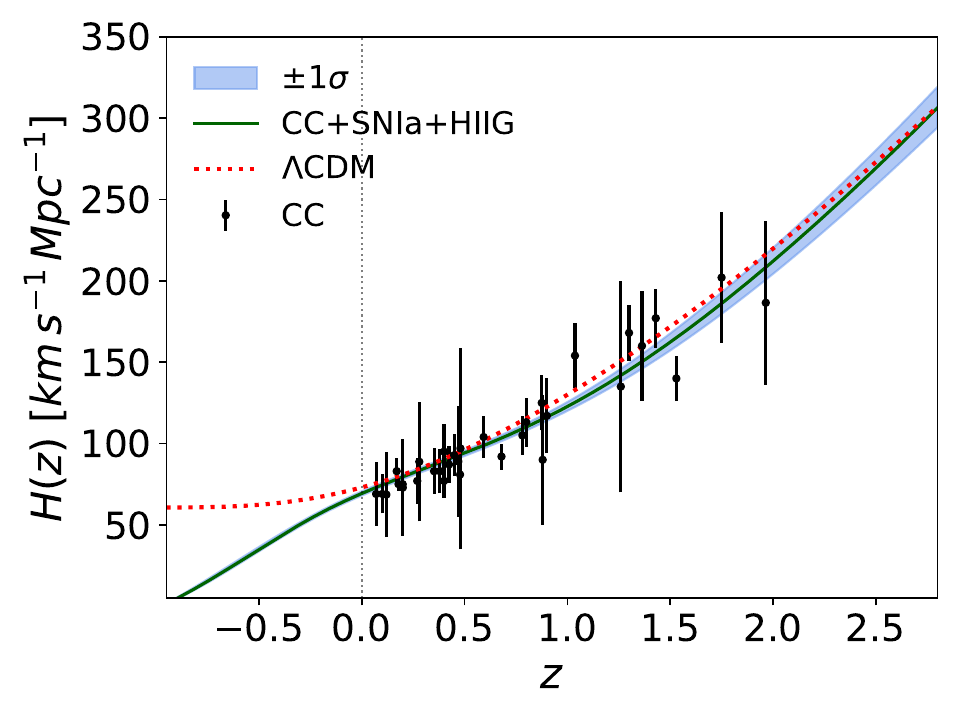}
    \includegraphics[width=0.31\textwidth]{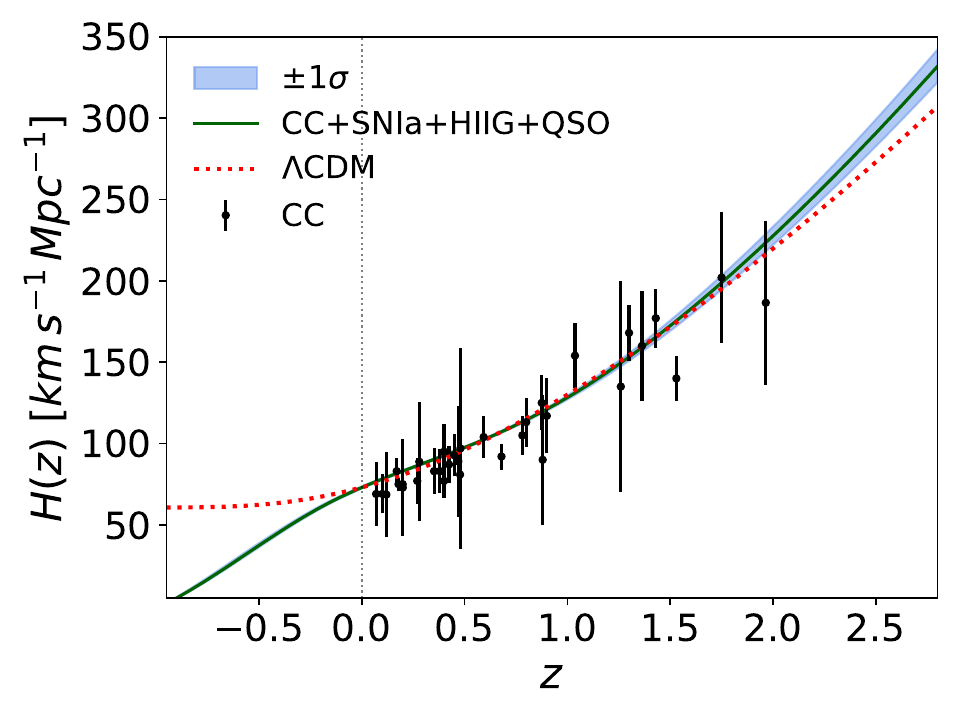}\\
    \includegraphics[width=0.31\textwidth]{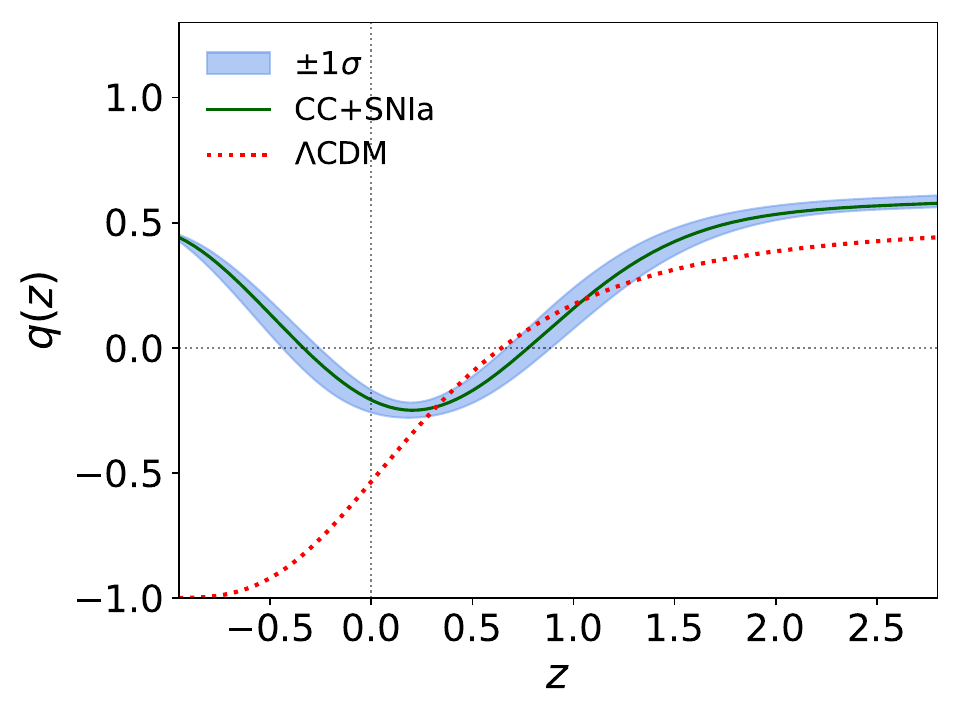}
    \includegraphics[width=0.31\textwidth]{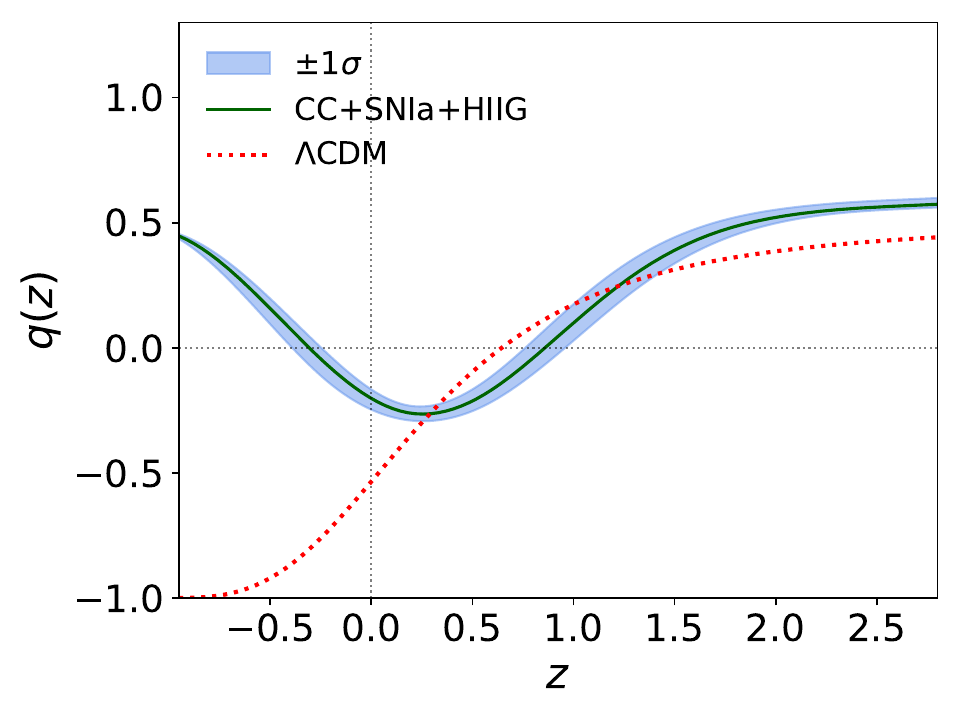}
    \includegraphics[width=0.31\textwidth]{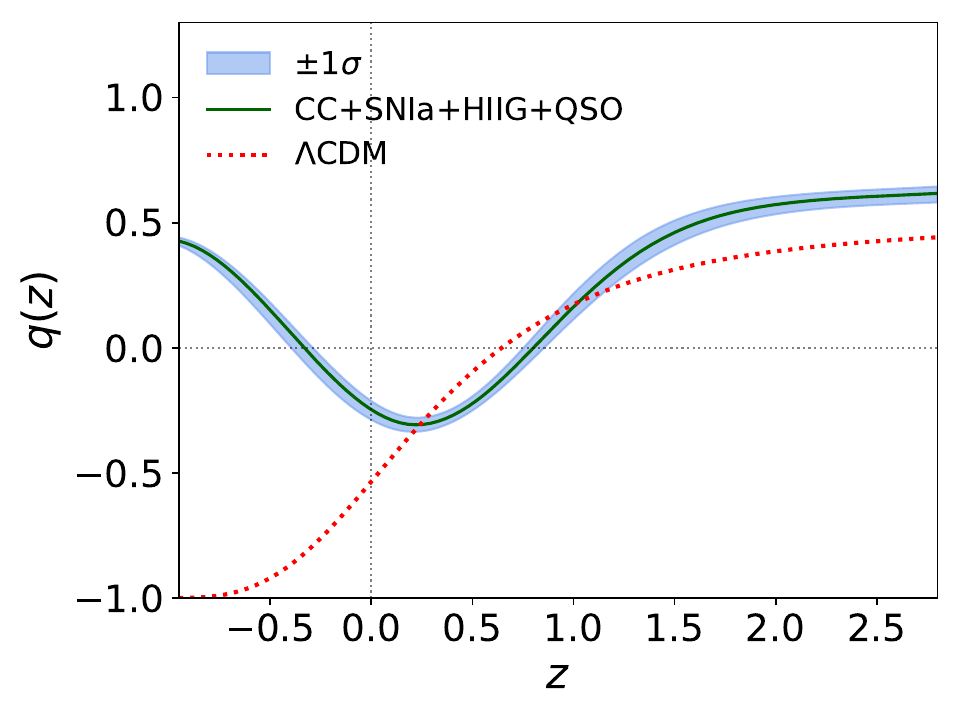}\\
    \includegraphics[width=0.31\textwidth]{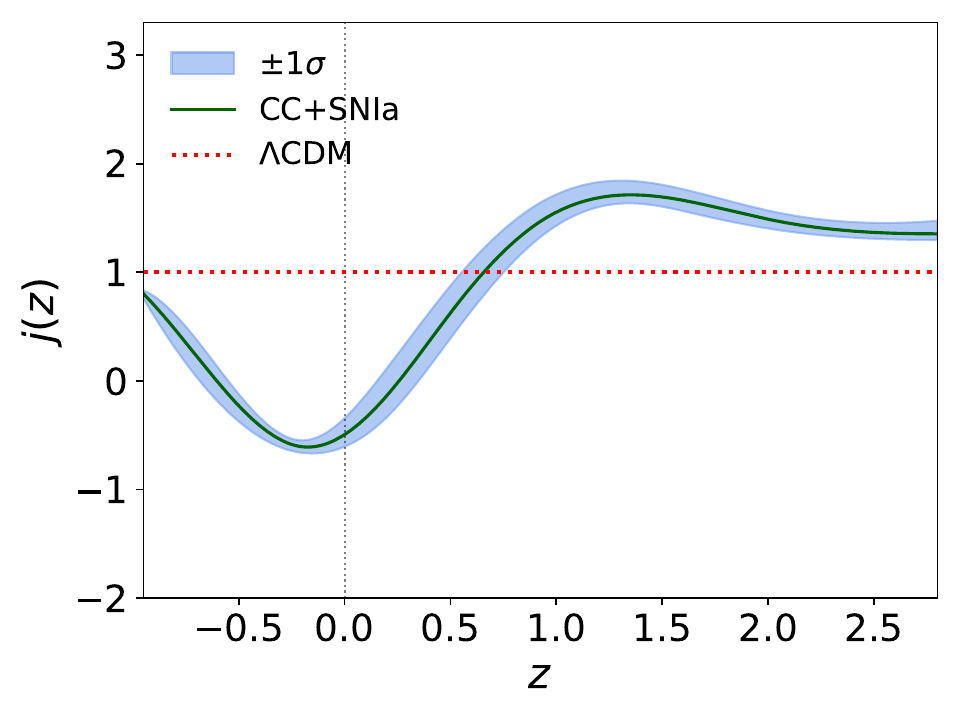}
    \includegraphics[width=0.31\textwidth]{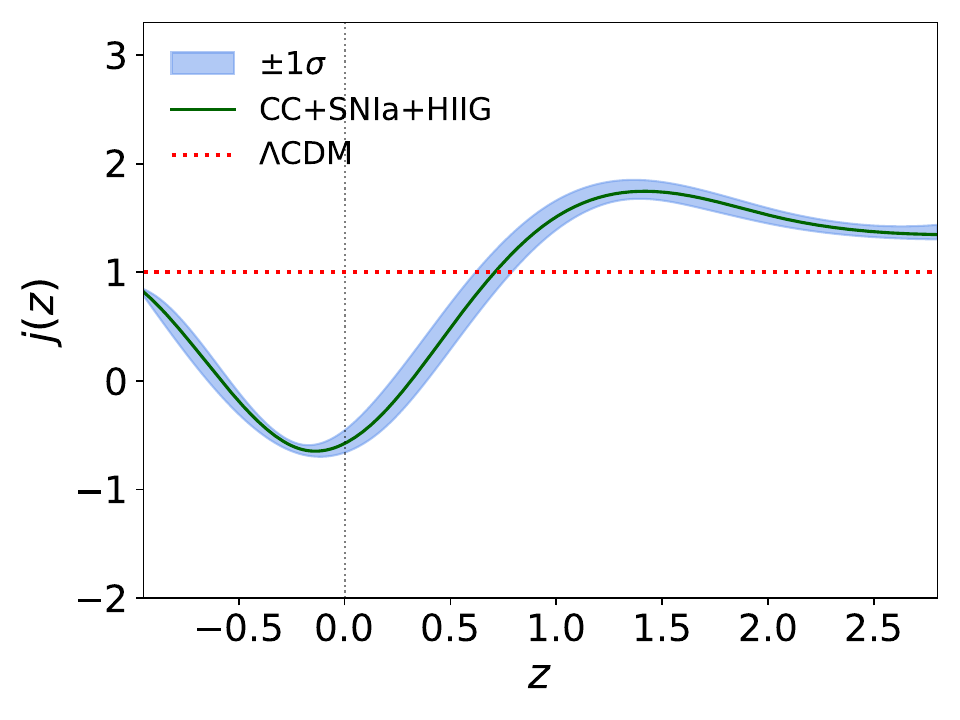}
    \includegraphics[width=0.31\textwidth]{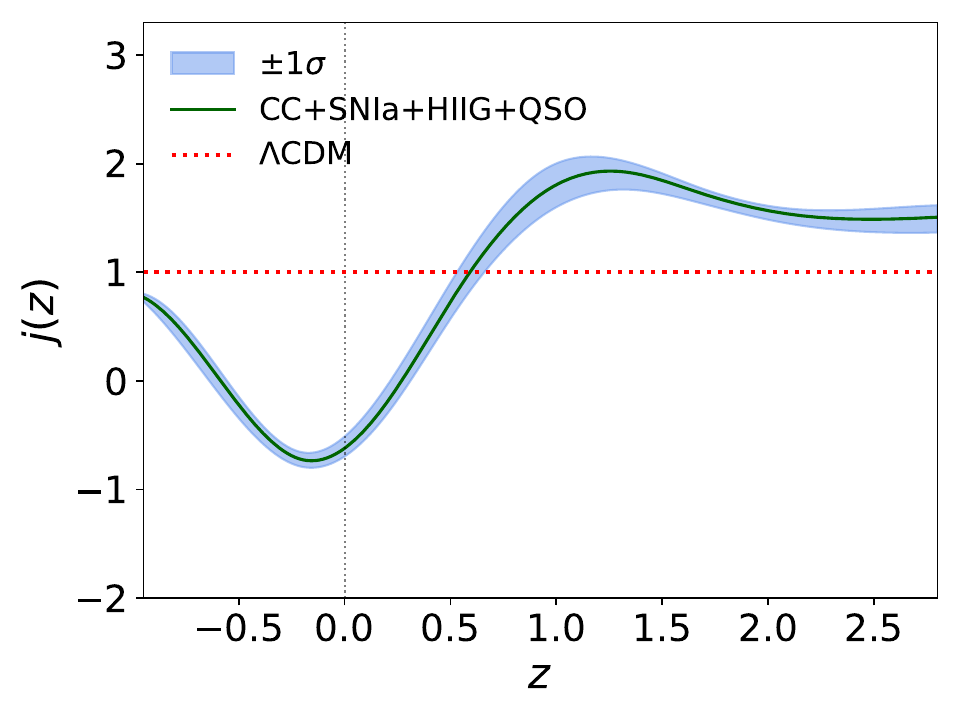}\\
    \includegraphics[width=0.31\textwidth]{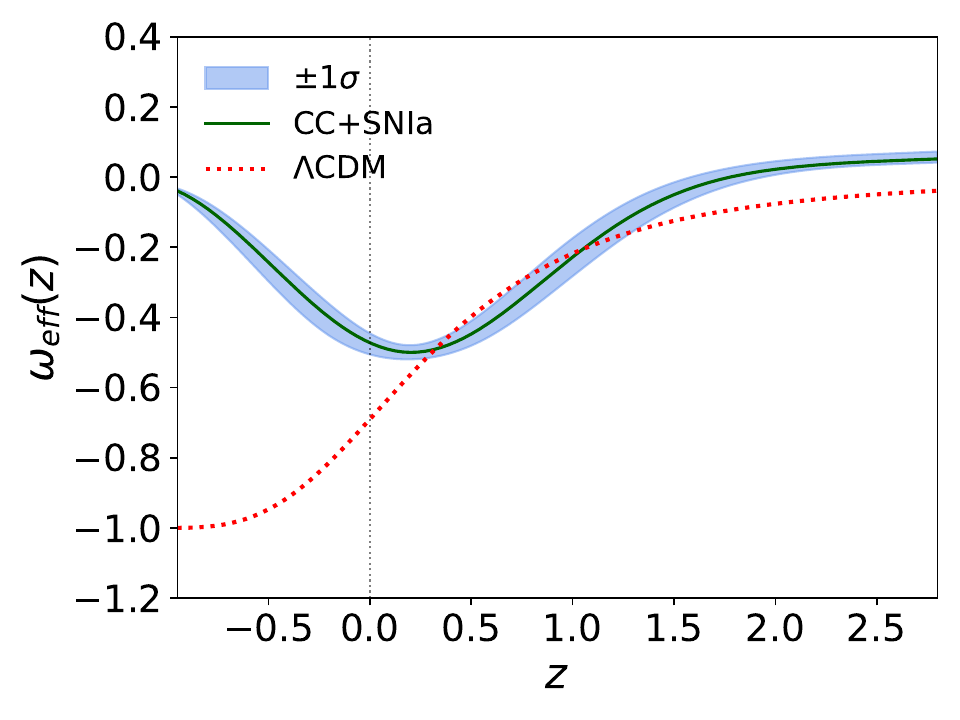}
    \includegraphics[width=0.31\textwidth]{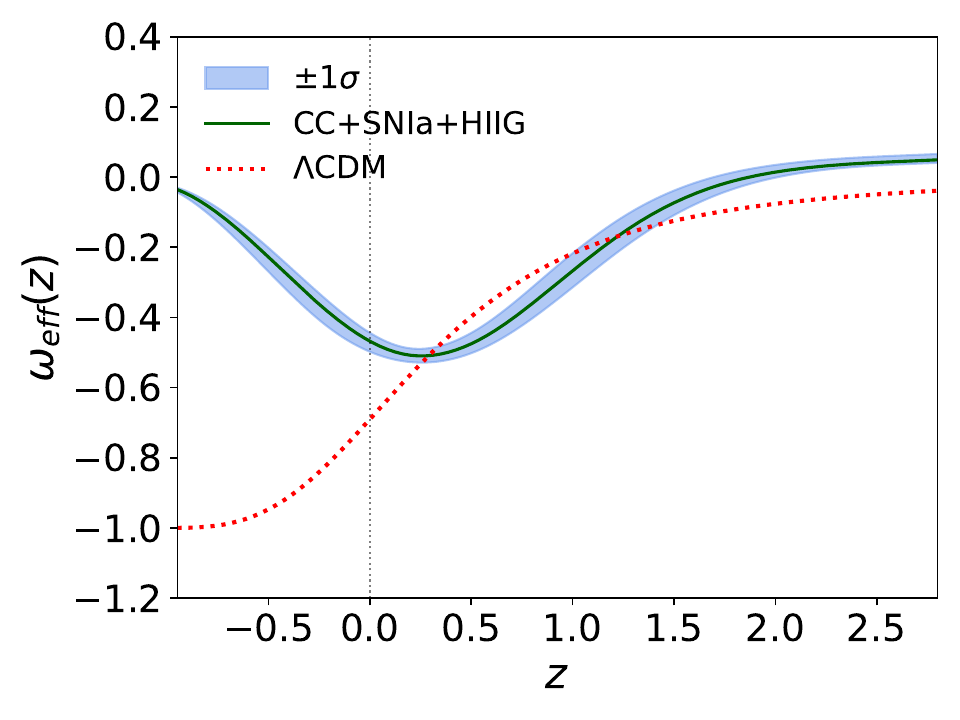}
    \includegraphics[width=0.31\textwidth]{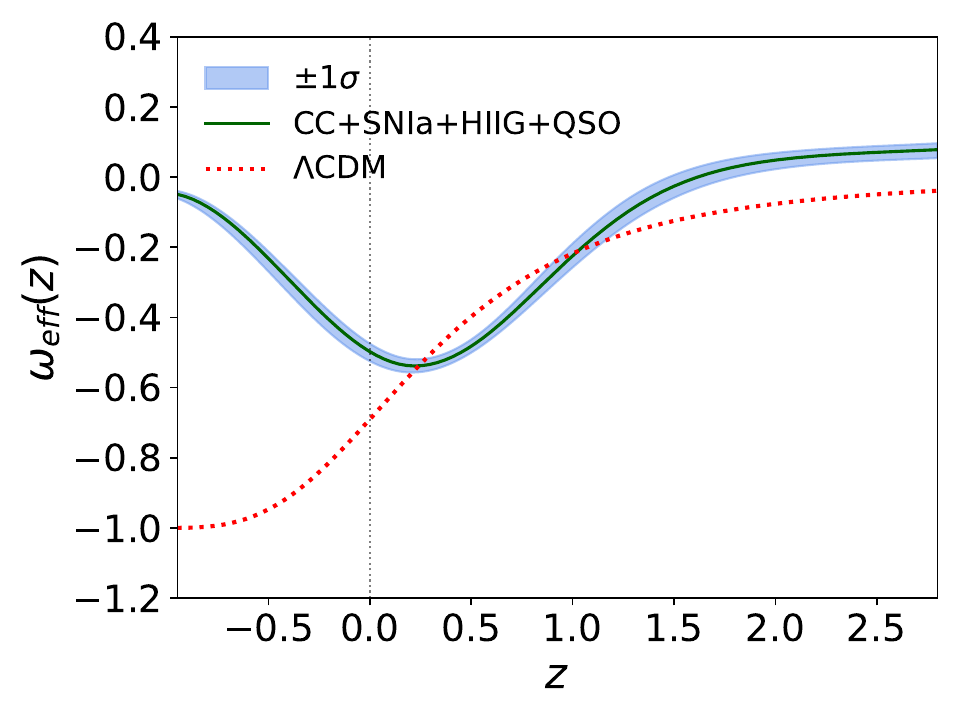}\\
    \caption{Reconstructions of the expansion history for our model parametrization  using three data combinations (left to right): CC+SNIa, CC+SNIa+HIIG, and CC+SNIa+HIIG+QSO. From top to bottom, we show the Hubble parameter $H(z)$, the deceleration parameter $q(z)$, the jerk parameter $j(z)$, and the effective equation of state $w_{\rm eff}(z)$. Solid curves correspond to the posterior median reconstruction, with shaded regions indicating the $1\sigma$  intervals. The reference $\Lambda$CDM prediction is overplotted as a red dashed curve. The negative-redshift domain ($z<0$) illustrates the model extrapolation into the near future.}
    \label{fig:cosmography}
\end{figure*}

\section{Observational Data}
\label{sec:ObsData}
The free parameter space is defined as ${\bf \Theta} = (h, q_0, z_e, z_c)$ and is constrained by assuming uniform priors within the ranges $0.2 < h < 1.0$, $-2.0 < q_0 < 0.5$, $0 < z_e < 20$, and $0 < z_c < 2$. The posterior distribution is explored using a Markov Chain Monte Carlo (MCMC) algorithm implemented through the \texttt{emcee} Python package \citep{Foreman:2013}. The convergence of the chains is assessed through the autocorrelation function, evaluated every 10 steps, and iterations are continued until the relative fluctuations in the estimated autocorrelation reach the order of $10^{-3}$. After this burn-in and convergence assessment phase, a total of 4000 chains, each with 200 steps, are generated to sample the parameter space $\Theta$ in the vicinity of the minimum of the negative log-likelihood function. This function is related to the statistic $\chi^2$ through   
$\log{\mathcal{L}} \propto -\chi^2/2$ 
and is constructed using the following datasets.

\begin{itemize}
\item Cosmic Chronometers (CC). We employ a dataset of 33 measurements of the Hubble parameter, $H(z)$, spanning the redshift range $0.07 < z < 1.965$ \cite{M_Moresco_2012, Moresco_2015, Moresco_2016, Moresco_2020, Jiao_2023, Tomasetti_2023}. This compilation is composed of 15 mutually correlated $H(z)$ measurements and 18 uncorrelated data points. Accordingly, the corresponding $\chi^2$ function is constructed as
\begin{equation}
    \chi^2_{\rm CC} = \sum_i^{18}\left( \frac{H_{\rm obs}^i - H_{\rm th}(z_i)}{\sigma^i} \right)^2 + \vec{H}\,{\rm Cov}^{-1}\vec{H}^{T},
\end{equation}
where the summation is performed over the uncorrelated subsample, while the second term accounts for the correlated measurements. Here, ${\rm Cov}^{-1}$ denotes the inverse of the covariance matrix associated with the vector $\vec{H}$ of correlated $H(z)$ data.

\item Type Ia supernovae (SNIa). The Pantheon+ compilation \cite{Scolnic2018-qf, Brout_2022} comprises 1701 measurements of the SNIa correlated distance modulus  that spans the redshift range $0.001 < z < 2.26$. We analyze these data using a correlated $\chi^2$ formalism \cite{Conley2010}, in which the nuisance parameters are analytically marginalized, yielding
\begin{equation}
    \chi^2_{\rm SNIa} = a + \log \left( \frac{e}{2\pi} \right) - \frac{b^2}{e},
\end{equation}
where 
\begin{eqnarray}
    a &=& \Delta\boldsymbol{\tilde{\mu}}^{T}\cdot\mathbf{Cov_{P}^{-1}}\cdot\Delta\boldsymbol{\tilde{\mu}}, \nonumber\\
    b &=& \Delta\boldsymbol{\tilde{\mu}}^{T}\cdot\mathbf{Cov_{P}^{-1}}\cdot\Delta\mathbf{1}, \\
    e &=& \Delta\mathbf{1}^{T}\cdot\mathbf{Cov_{P}^{-1}}\cdot\Delta\mathbf{1}\, , \nonumber
\end{eqnarray}
and $\Delta\boldsymbol{\tilde{\mu}}$ denotes the vector of residuals between the theoretical and observed distance moduli. The theoretical distance modulus is defined as
    \begin{equation}
\mu_{th}(z, \Theta) = 5 \log_{10} \left [ \frac{d_L(z)}{1\,{\rm Mpc}}\right] + 25,
\end{equation}
and is related to the luminosity distance $d_L$ via
    \begin{equation}\label{eq:dL}
d_L(z)=(1+z)c\int_0^z\frac{dz^{\prime}}{H(z^{\prime})}\,,
\end{equation}
where $c$ is the speed of light. Furthermore, $\Delta\mathbf{1}=(1,1,\dots,1)^T$ corresponds to the transpose of the unit vector and $\mathbf{Cov_{P}}$ denotes the Pantheon+ covariance matrix.

\item Hydrogen II galaxies (HIIG). The dataset consists of 181 distance-modulus measurements for compact star-forming systems with stellar masses $M < 10^9 M_{\odot}$, classified as HIIG, spanning the redshift interval $0.01 < z < 2.6$ \cite{GonzalezMoran2019, Gonzalez-Moran:2021drc}. The corresponding statistic $\chi^2$ is defined as
\begin{equation}\label{eq:chi2_HIIG}
   \chi^2_{\rm HIIG} = \sum_{i=1}^{181} \frac{\left[\mu_{\rm th}(z_i, \Theta) - \mu_{\rm obs}^i\right]^2}{\epsilon_i^2},
\end{equation}
where $\mu_{\rm obs}^i \pm \epsilon_i$ denotes the empirically determined distance modulus and its associated uncertainty at redshift $z_i$, and $\mu_{\rm th}(z_i, \Theta)$ is the corresponding theoretical prediction for a given set of cosmological parameters $\Theta$.
    
\item Intermediate-luminosity quasars (QSO). We employ a sample of 120 angular-size measurements of ultra-compact radio quasars spanning the redshift range $0.462 < z < 2.73$ \cite{ShuoQSO:2017}. Due to their negligible dependence on both redshift and intrinsic luminosity, these sources can be modeled as standard rulers characterized by a fixed comoving linear size $l_m$. The observed angular size $\theta$ is related to the angular diameter distance $D_A(z)$ and the intrinsic linear size $l_m$ through
\begin{equation}
\theta(z) = \frac{l_m}{D_A(z)}\,.
\end{equation}
In this analysis, we adopt $l_m = 11.03 \pm 0.25~\mathrm{pc}$, as calibrated in Ref.~\cite{ShuoQSO:2017}, which corresponds to a characteristic radial distance at which active galactic nucleus jets become optically thick at the observed frequency of approximately 2 GHz. The corresponding contribution to total likelihood is quantified by an uncorrelated statistic $\chi^2$,
\begin{equation}
\chi^2_{\rm QSO} = \sum_{i=1}^{120} \left( \frac{\theta_{\mathrm{obs}}(z_i) - \theta_{\mathrm{th}}(z_i)}{\sigma_{\theta_{\mathrm{obs}}}(z_i)} \right)^2\,,
\end{equation}
where $\theta_{\mathrm{obs}}(z_i) \pm \sigma_{\theta_{\mathrm{obs}}}(z_i)$ denotes the measured angular size and its associated uncertainty $1\sigma$ at redshift $z_i$, and $\theta_{\mathrm{th}}(z_i)$ is the corresponding theoretical prediction.
\end{itemize}

As a baseline dataset, we use the combination CC+SNIa and then CC+SNIa+HIIG and CC+SNIa+HIIG+QSO.

\begin{figure*}
    \centering
    \includegraphics[width=0.8\textwidth]{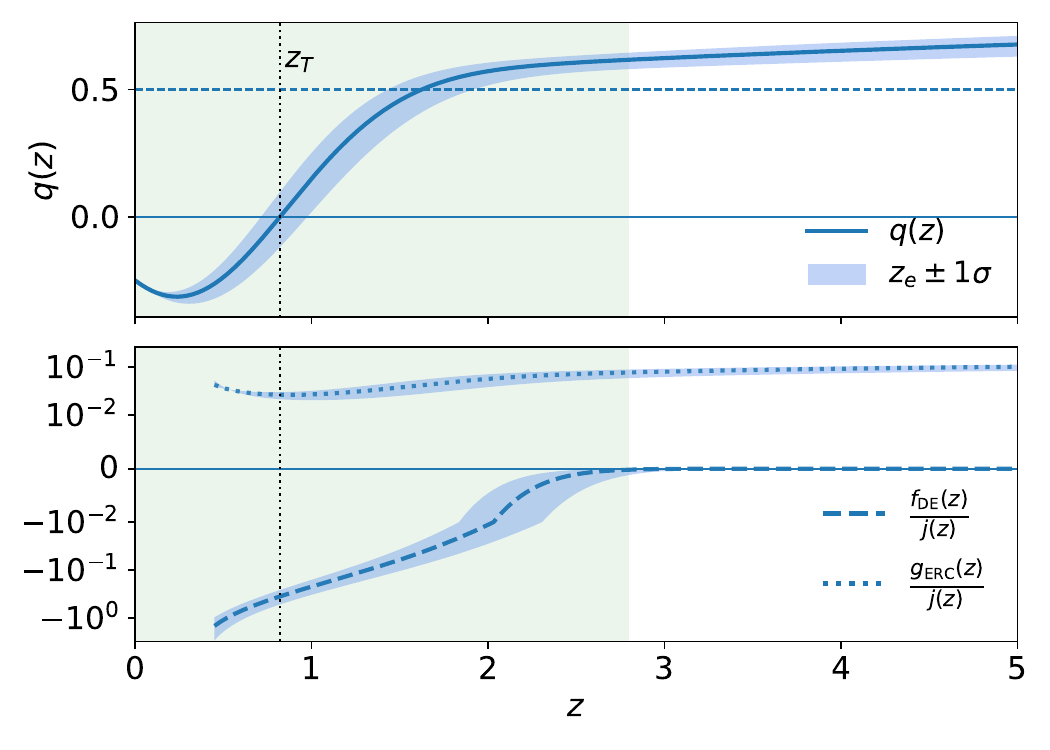}
    \caption{\textit{Upper panel:}  reconstructed $q(z)$ for the median best-fit parameters (solid curve). 
The shaded band represents the $1\sigma$ variation induced by the scale parameter $z_e$, while keeping the remaining parameters fixed at their median values. The vertical  green region marks the redshift interval directly constrained by the data. \textit{Lower panel:} ratios $f_{\mathrm{DE}}(z)/j(z)$ (dashed) and $g_{\mathrm{ERC}}(z)/j(z)$ (dotted), normalized by the reconstructed jerk parameter $j(z)$. }
    \label{fig:fgJerk}
\end{figure*}

\section{Results  and Discussions} \label{sec:Results}

\subsection{Late-time constraints on a decelerating Universe}

\label{subsec:decelerating_universe}

Table~\ref{tab:bf_model} summarizes the median values and $1\sigma$ confidence intervals of the free parameters obtained from the MCMC analysis using the different combinations of late-time observables, namely CC, SNIa, HIIG, and QSO. In addition to the parameters $q_0$, $z_c$, $z_e$, and the Hubble parameter $h$, the table includes the inferred cosmic age $\tau_U$ and the redshift $z_T$ at which the Universe transitions from a decelerated to an accelerated phase. Note that $z_c$ does not coincide with $z_T$. While $z_c$ controls the location of the  modulation in the dark-energy contribution $f_{\rm DE}(z)$, the actual transition redshift $z_T$ is defined implicitly by the condition $q(z_T)=0$. Before discussing the implications of these results, it is worth noting that the reconstruction of $q(z)$ at higher redshifts is largely driven by the assumed functional form, since current observational data mainly constrain the low-redshift regime.

For the complete combination of data (last column in Table~\ref{tab:bf_model}), the present-day deceleration parameter is found
to be  $q_0 = -0.25$, consistent with a late-time accelerated expansion, while
the transition redshift is very well constrained around $z_T\simeq0.8$, consistent with values reported in the literature \citep[see][and references therein]{2016IJMPD..2550032A,RomanGarza2019,2023RAA....23i5001D}. On the other hand, the characteristic scale $z_e$ remains weakly constrained by data, since its role is to regulate the asymptotic high-redshift behavior of the expansion rather than to control the late-time transition. This behavior reflects the fact that the observations used in this paper are mainly sensitive to the low- and intermediate-redshift evolution of $q(z)$, while providing limited constraints on its functional form at high redshifts. The physical implications of this feature and its connection to the ERC are discussed in detail in the following section.

Figure~\ref{fig:contours} displays the  marginalized posterior
distributions and the corresponding two-dimensional confidence regions for the main model parameters.  These constraints highlight the ability of the parametrization to probe the redshift epoch during which the expansion history deviates from matter domination. In particular, the reconstructed deceleration parameter $q(z)$ shows a characteristic trend over the range probed by the data, departing from the
$\Lambda$CDM expectation and showing a tendency towards reduced acceleration at late times. This behavior is consistently reflected in the reconstructed effective
equation of state $w_{\rm eff}(z)$, as shown in Fig.~2. Such features are in  agreement with recent late-time analyzes based on DESI DR2 measurements, which also report evidence for a nontrivial evolution of the deceleration parameter \cite{2025PhRvD.112h3511L, 2025NatAs...9.1879G}. Although the reconstructed behavior of $q(z)$ suggests a tendency toward a reduced late-time acceleration compared to the standard $\Lambda$CDM scenario, this feature should be interpreted with caution. Its statistical significance is limited and depends on the adopted parametrization.

 It is instructive to compare our kinematic transition redshift ($z_T \simeq 0.8$) with recent results obtained in modified gravity 
frameworks. For instance, studies in $f(R, L_m)$ gravity \cite{10.1016/j.dark.2024.101545} 
and $f(Q, T)$ gravity \cite{10.1142/S0219887824500543} report transition redshifts 
in the range $z_T \in [0.6, 0.8]$, which are broadly consistent with our 
reconstruction. In this context, it is worth noting that these approaches typically embed the 
expansion history within specific gravitational frameworks, while our analysis 
is based on a phenomenological parametrization constrained directly by late-time 
observational data. The overall agreement in $z_T$ suggests that the transition 
redshift is a relatively robust feature across different modeling strategies, 
while the flexibility of the present approach allows us to explore possible 
deviations from the standard $\Lambda$CDM expectation within the redshift range 
effectively probed by the data.

The tight constraints on $z_T$ are related to the sensitivity of both distance indicators (SNIa, HIIG, and QSOs) and expansion-rate measurements from cosmic chronometers, which probe the local evolution of $H(z)$,  to the transition in sign of $q(z)$,  which is directly encoded in the cosmic jerk through the exact kinematic relation between $q(z)$ and $j(z)$ \cite{2005ASPC..339...27B, 2004CQGra..21.2603V}.  Conversely, the wider posterior distribution of $z_e$ indicates that it functions as an asymptotic regulator at high redshifts, having only a secondary influence over late-time cosmic dynamics. As illustrated in Fig.~\ref{fig:cosmography}, the reconstructed jerk exhibits a clear deviation from its $\Lambda$CDM value around the transition epoch, providing an efficient kinematic tracer of the transition to the accelerated expansion phase\cite{2003JETPL..77..201S}.

It is important to note that the inferred cosmic age $\tau_U$ is a derived, model-dependent quantity obtained from the kinematic reconstruction constrained by low and intermediate redshift data only and thus should be interpreted with caution. In the absence of early probes such as the CMB, the resulting age is sensitive to the inferred value of $H_0$ and to the specific form of $q(z)$ at late times. The low value of $\tau_U$ reported in Table~1 does not represent a direct measurement of the absolute age of the Universe, consistent with current estimates of  $t_0 \simeq 13.8~\mathrm{Gyr}$ \cite{2013ApJS..208...20B,Planck:2018, 2020JCAP...12..047A}. This shift is naturally associated with the relatively large value of $h$ preferred by our late-time reconstruction, since for a broadly similar expansion history a higher $H_0$ directly translates into a lower inferred age. We emphasize that no early-time data is used in the present analysis. A more robust determination of the cosmic age within this framework would require the inclusion of high-redshift information, which is beyond the scope of the present work and will be addressed in a future study. Recent analyses suggest that an oldest age significantly below $\sim 13\,\mathrm{Gyr}$ would be difficult to reconcile with stellar population data  \cite[see, e.g.][]{2025arXiv250902692T}). Although our inferred value should not be interpreted as a direct constraint on the absolute cosmic age, this comparison highlights the importance of incorporating early-Universe information in future extensions of the model. This reinforces the fact that the inferred cosmic age should be interpreted within the limitations of the adopted parametrization.

\begin{figure*}
    \centering
    \includegraphics[width=0.8\textwidth]{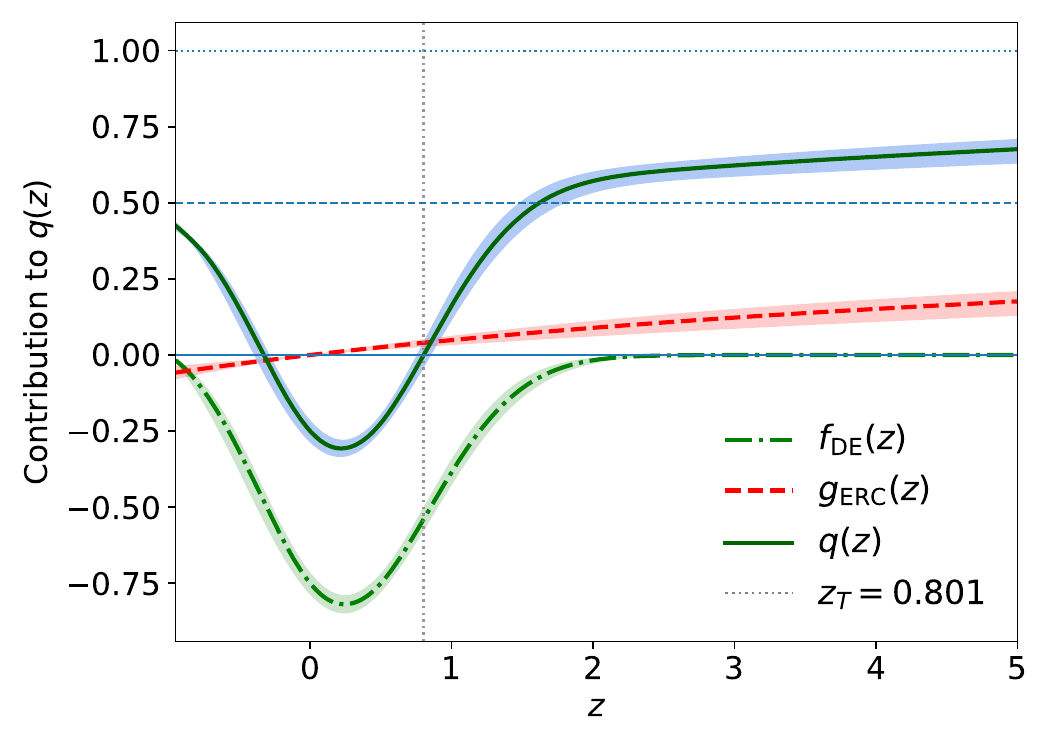}
\caption{Decomposition of the reconstructed deceleration parameter $q(z) = 1/2 + f_{\mathrm{DE}}(z) + g_{\mathrm{ERC}}(z)$ 
(see Eq.~\eqref{eq:q_split}) as a function of redshift.  The dashed–dotted curve shows the late-time contribution $f_{\mathrm{DE}}(z)$, the dashed curve corresponds to the 
effective radiative component $g_{\mathrm{ERC}}(z)$, and the  solid line represents the total $q(z)$. Shaded regions indicate the $1\sigma$ confidence intervals  
The vertical dotted line marks the transition redshift $z_T$.
}

    \label{fig:contribution_qz}
\end{figure*}

\subsection{Kinematic diagnostics beyond the data range}
\label{subsec:Kinematic}

Existing efforts have explored dark-energy dynamics across pre- and post-recombination epochs by introducing  parametrizations capable of capturing distinct behaviors in different cosmic eras \cite[e.g.,][]{2021PhRvD.104h3536G}. In contrast, the reconstruction discussed in the previous subsection is constrained exclusively by
low- and intermediate-redshift observations (with the available data extending up to
$z \simeq 2.8$), and the adopted parameterization of $q(z)$ allows us to explore, in a controlled  way, the behavior of the expansion history beyond the redshift range directly probed by the data.

Figure~\ref{fig:fgJerk} presents a kinematic diagnostic of the reconstructed
deceleration parameter for $z \gtrsim 0.4$. The upper panel shows the reconstructed
$q(z)$ for the median parameters together with representative variations of the scale $z_e$, which regulates the asymptotic behavior at high redshift. Although all curves
remain fully consistent within the observation window (shadow region), they gradually separate at higher redshifts, illustrating how $z_e$ governs the extrapolation of the expansion
history while preserving the late-time transition dynamics. The lower panel displays the ratios $f_{\mathrm{DE}}(z)/j(z)$ and $g_{\mathrm{ERC}}(z)/j(z)$, using the reconstructed jerk parameter $j(z)$. This normalization is motivated by the fact that the jerk parameter controls the redshift evolution of the deceleration parameter and thus captures deviations from $\Lambda$CDM kinematics \cite{2016PhRvD..93d3002M}. Expressing the dynamical functions relative to 
$j(z)$ allows us to quantify the relative strength of dark-sector effects against the intrinsic evolution of the expansion rate. The figure suggests an interplay between the kinematic roles of the
different components: the late-time contribution $f_{\mathrm{DE}}(z)$ dominates around the transition epoch and rapidly decays at higher redshifts, while the effective radiative component $g_{\mathrm{ERC}}(z)$ remains subdominant at low redshift but becomes increasingly important as $z$ grows, contributing to a smooth and well-behaved asymptotic behavior.

Note that this analysis should not be seen as a reconstruction of $q(z)$ at high redshift. Since no early-time data are included in the fit, the behavior beyond the data range reflects the internal kinematic structure of the parametrization rather than additional observational constraints. Although cosmographic extensions to higher redshift using GRB data have been explored in the literature \cite[e.g.,][]{2012MNRAS.426.1396D}, such approaches remain sensitive to convergence effects. In this sense, the figure provides a consistency check of the extrapolation implied by the model, indicating that the late-time reconstruction of $q(z)$  remains stable while the high-redshift behavior is naturally regulated by $z_e$. A physical interpretation of this parameter and its connection with the ERC are discussed in detail in the following subsection.

\begin{figure*}
    \centering
    \includegraphics[width=0.8\textwidth]{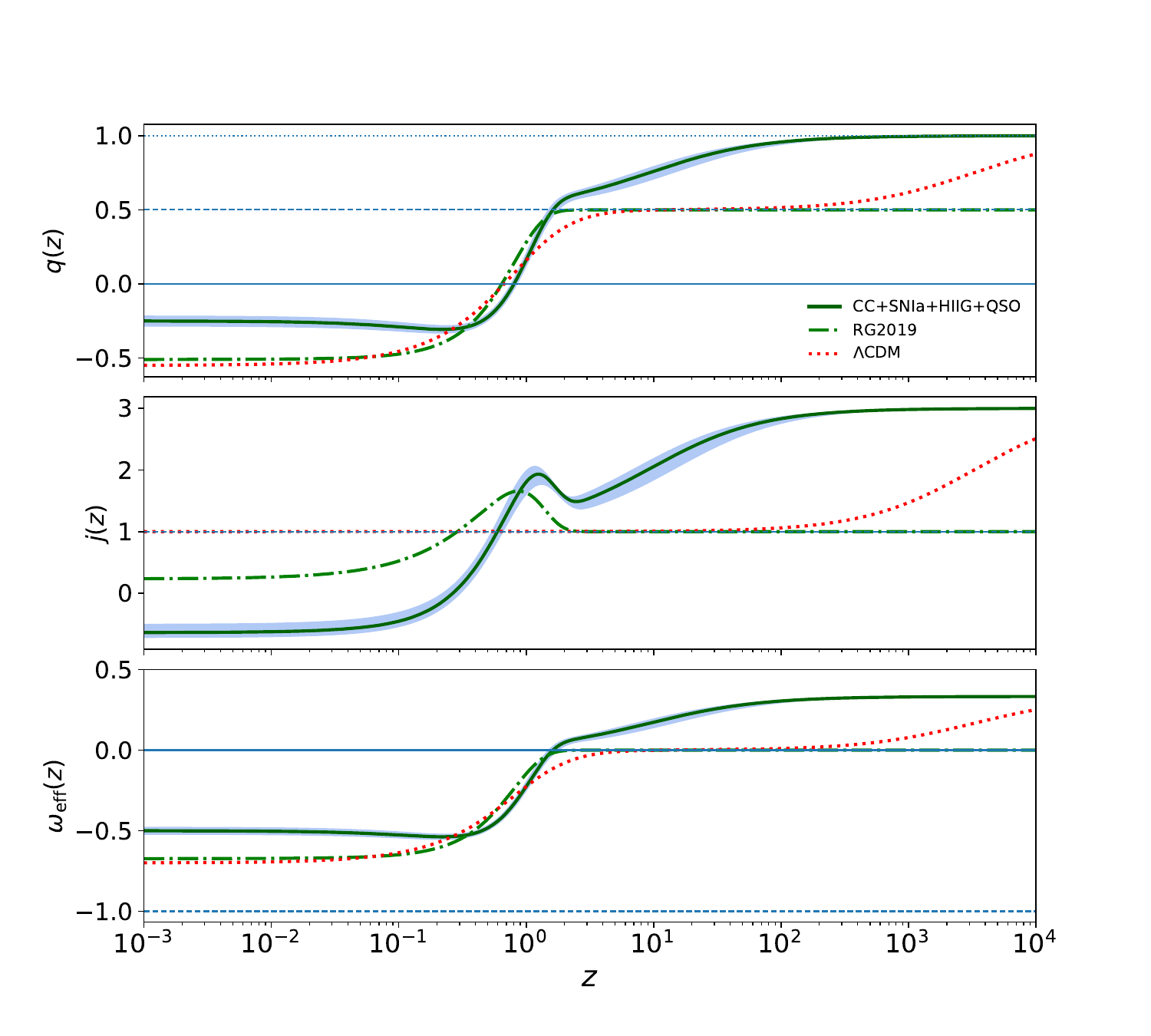}
\caption{Extended evolution of the kinematic quantities $q(z)$, $j(z)$ and 
$\omega_{\rm eff}(z)$ over the interval $10^{-3} < z < 10^{4}$. The solid dark-green curves correspond to the reconstruction obtained from CC+SNIa+HIIG+QSO data, with shaded bands indicating the 
$1\sigma$ uncertainties. For comparison, the RG2019 parametrization 
(dash–dotted green) and the $\Lambda$CDM model (red dotted) are also shown. 
The proposed model exhibits a smooth and monotonic behavior across the 
full redshift range.}
    \label{fig:qz_jz_weff}
\end{figure*}

\subsection{The role of the ERC}
\label{subsec:erc}

Figure~\ref{fig:contribution_qz} shows how the deceleration parameter is built from the two dynamical functions that define Eq.~(1). Contribution $f_{\mathrm{DE}}(z)$ dominates the dynamics around the transition epoch, driving late-time acceleration and controlling the shape of $q(z)$ near $z_T$. In contrast, the effective radiative contribution $g_{\mathrm{ERC}}(z)$ remains subdominant at low redshift but increases smoothly with $z$, becoming progressively relevant as the Universe approaches the matter-dominated regime. 

The ERC does not act as a driver of acceleration; instead, it stabilizes the high-redshift behavior of the parametrization, regulating the asymptotic limit without altering the late-time transition dynamics. This can be seen from the fact that $g_{\mathrm{ERC}}(z)$ modifies the slope of $q(z)$ at intermediate redshifts while preserving both the transition's location and the reconstructed low-$z$ constraints. The scale parameter $z_e$  can therefore be interpreted as an effective high-redshift cutoff rather than as a tightly constrained physical quantity. Given the absence of early-Universe data in the present analysis, the posterior distribution remains broad and partially prior-driven. Within this framework, values in the approximate range $8 \lesssim z_e \lesssim 14$ are compatible with the late-time constraints (see Table~\ref{tab:bf_model}). This interval overlaps with the redshift commonly associated with the epoch of reionization \cite{2006AJ....132..117F,Planck:2018}. Although the ERC is introduced purely at the kinematic level and does not represent a physical radiation component, the coincidence of scales suggests that the parametrization naturally transitions toward a behavior consistent with the expected matter-dominated regime near epochs where the expansion history undergoes a fundamental shift in its dynamic regime and represents an important era in the evolution of the
Universe \cite{2015ApJ...802L..19R}.

The stability of the proposed model is  illustrated in Fig.~\ref{fig:qz_jz_weff}, which displays the evolution of the kinematic parameters $q(z)$, $j(z)$, and $w_{\rm eff}(z)$ over a wide redshift range ($10^{-3} < z < 10^4$). For comparison, we include the $\Lambda$CDM baseline and the reconstruction from \cite{RomanGarza2019}.  While the late-time acceleration is dominated by the $f_{\mathrm{DE}}(z)$ term, the introduction of the ERC produces a well-behaved transition toward the matter-dominated regime. This is particularly evident in the $q(z)$ panel, where our model (dark-green solid line) approaches the expected decelerating plateau ($q \approx 0.5$) at intermediate redshifts and smoothly approaches the radiation-dominated limit ($q \rightarrow 1$) at very high $z$. Unlike the parameterization of \cite{RomanGarza2019}, which remains fixed at $q=0.5$ for $z > 2$, our model captures the emerging influence of the effective radiative contribution.

Furthermore, the jerk parameter $j(z)$ and the equation of state $w_{\rm eff}(z)$ exhibit a smooth and continuous evolution over the entire redshift interval. No abrupt features or oscillations arise from the parametrization. This behavior is consistent with the internal structure of the model and the stabilizing role of the ERC contribution at intermediate and high redshifts. In addition, the jerk parameter is not assumed a priori \cite{2013PhLB..727....8Z}; instead, it is derived consistently from the reconstructed deceleration history through its kinematic definition \cite{2018EPJC...78..862A}. As a result, the evolution of $j(z)$ directly reflects the internal structure of the $q(z)$ parametrization rather than an independent phenomenological ansatz.

\section{Conclusions} \label{sec:Conclusions}

This paper is devoted to presenting a generalized phenomenological parameterization of the deceleration parameter $q(z)$ for reconstructing the late-time expansion history in a purely kinematic and model independent framework. The proposed form extends the two--parameter reconstruction of \cite{RomanGarza2019} by introducing an ERC, designed to regulate the asymptotic high redshift behavior while preserving the localized late-time transition to acceleration. Our findings reinforce the importance of employing flexible parametrizations 
capable of describing the expansion history without relying on a specific underlying gravitational framework. Recent studies have explored related late-time behaviors within modified gravity and effective-fluid frameworks, including observationally constrained 
 $f(T)$-based parametrizations \cite{10.1142/S0219887824501664} and phenomenological $f(T, T_G)$-inspired effective-fluid analyses \cite{10.1142/S0219887826501409}. In this context, our framework provides a complementary perspective on the reconstructed dynamics. In particular, the controlled high-redshift behavior, 
regulated by the parametrization, is consistent with a smooth transition toward 
the matter-dominated regime, while maintaining agreement with late-time 
observational constraints.

Using late-time observational data from cosmic chronometers, Pantheon+ Type Ia supernovae, HII galaxies, and intermediate-luminosity quasars, we constrained the parameter set $h$, $q_0$, $z_c$, $z_e$ through a standard Bayesian MCMC analysis. For the full data combination, we obtain a  deceleration parameter $q_0 \simeq -0.25$, consistent with a currently accelerating Universe, and a well constrained transition redshift $z_T \simeq 0.8$, suggesting that the onset of acceleration may occur earlier than in the $\Lambda$CDM model. The Hubble parameter is reconstructed with a relatively high value, $h \simeq 0.729$, consistent with local  determinations such as the SH0ES measurement $H_0 = 73.04 \pm 1.04 \ \mathrm{km\,s^{-1}\,Mpc^{-1}}$ \citep{Riess_2022}, and recent measurements using JWST\citep{ 2025ApJ...992L..34R}. In this sense, our late time kinematic reconstruction  lies within the observational regime commonly associated with the Hubble tension. The early time scale parameter $z_e$ remains weakly constrained, reflecting the fact that the data used here mainly probe the low- and intermediate-redshift evolution of the expansion history.

As shown in Fig.~\ref{fig:cosmography}, the reconstructed expansion history exhibits a tendency toward a reduced late-time acceleration compared to $\Lambda$CDM. Although the present-day value $q_0$ remains negative, indicating an accelerating Universe, the deviation from the standard model suggests that the acceleration rate may be weaker than in the concordance scenario. In particular, the extrapolated behavior of $H(z)$ toward the future limit $z \to -1$ remains regular and approaches a well-behaved asymptotic state. Although this asymptotic property is not exclusive to the present parametrization \citep[see, e.g.,][]{2003RvMP...75..559P, 2003PhRvL..90i1301L}, the reconstructed kinematic evolution highlights a smoother and less rapidly accelerating late-time regime relative to $\Lambda$CDM. This result is particularly relevant in light of the ongoing debate on the possible weakening of cosmic acceleration at very low redshift. Several recent  analyses based on late-time probes have reported indications of dynamical dark energy behavior and potential departures from a strictly constant cosmological constant, including trends compatible with a reduced acceleration rate or evolving equation of state \citep{DESI:2025zgx, 2025PhRvD.112h3511L, 2025NatAs...9.1879G}. Although our reconstruction does not imply a present-day decelerating Universe, it remains fully consistent with scenarios in which the acceleration is diminishing relative to the $\Lambda$CDM expectation.

The behavior of the reconstructed $q(z)$ is consistently reflected in the derived effective equation of state, $w_{\rm eff}(z)$, and in the evolution of the jerk parameter, $j(z)$. Importantly, the jerk parameter is not imposed as an independent ansatz but is derived self-consistently from the reconstructed deceleration history through its kinematic definition. Consequently, its evolution is fully determined by the functional form of $q(z)$ and its redshift derivative, so that no additional phenomenological assumptions are introduced in $j(z)$.  In this sense, the purpose of introducing the ERC is not to improve the late-time fit but to provide a controlled and well-behaved extension of the deceleration parameter beyond the redshift range directly probed by the data. In the observation window considered in this work, the contribution of the ERC is naturally subdominant and does not significantly affect the  reconstruction,  recovering the behavior of \cite{RomanGarza2019} at low redshift while ensuring a smooth transition toward the expected asymptotic limits at higher redshift.

We emphasize that the estimate of cosmic age should be interpreted with caution in a late-time-only reconstruction, as it relies on the inferred value of $H_0$ and on the extrapolation of $q(z)$ beyond the redshift range directly constrained by the data. Incorporating additional constraints from the measurements of BAO~\cite{DESI:2025zgx} and CMB~\cite{Planck:2018} within this kinematic framework constitutes a natural next step and will allow a more stringent assessment of the model at high redshift.

\backmatter


%


\bmhead{Acknowledgements}
We thank anonymous referee for thoughtful remarks and suggestions.
The authors acknowledge support from the Secretar\'ia de Ciencias, Humanidades, Tecnolog\'ia e Innovaci\'on (SECIHTI) under grant CBF-2025-I-551 (“Convocatoria Ciencia B\'asica y de Frontera 2025”) and from project ANID Vinculaci\'on Internacional FOVI240098. A.H.A. thanks to the support from Luis Aguilar, 
Alejandro de Le\'on, Carlos Flores, and Jair Garc\'ia of the Laboratorio 
Nacional de Visualizaci\'on Cient\'ifica Avanzada. M.A.G.-A. acknowledges support from c\'atedra Marcos Moshinsky, SECIHTI for the support with the National Research System (SNII) grant and the project 0056 from Universidad Iberoamericana: Nuestro Universo en Aceleraci\'on, energ\'ia oscura o modificaciones a la relatividad general. The numerical analysis was also carried out by {\it Numerical Integration for Cosmological Theory and Experiments in High-energy Astrophysics} (Nicte Ha) cluster at IBERO University, acquired through c\'atedra MM support. A.H.A and M.A.G.-A  acknowledge partial support from project ANID Vinculaci\'on Internacional FOVI220144. V.M. acknowledges partial support from Centro de Astrof\'{\i}sica de Valpara\'{\i}so CIDI 21.
JM acknowledges partial support from project Fondecyt Regular No. 1240514.

\section*{Declarations}


\begin{itemize}
\item Availability of data and materials

This manuscript has no associated data or the data will not be deposited. [Authors’ comment: This is theoretical research work, and is based upon the analysis of public observational data (previously published by different authors). So no additional data are associated with this work.]

\end{itemize}







\begin{appendices}

\section{Motivation of the functional form of $q(z)$}
\label{app:motivation}

We briefly discuss the kinematic motivation underlying the
functional form adopted for the deceleration parameter $q(z)$ in
Equation~\eqref{eq:q_final}. The purpose of this discussion is not to provide a fundamental
derivation but rather to clarify the rationale behind the chosen
parametrization.

The deceleration and jerk parameters are defined as
\begin{equation}
q \equiv -\frac{\ddot a}{aH^2},
\qquad
j \equiv \frac{\dddot a}{aH^3}.
\end{equation}
From these definitions, one obtains the exact identity
\begin{equation}
j(z) = q(z)\bigl[2q(z)+1\bigr] + (1+z)\frac{dq}{dz},
\label{eq:jerk_identity_app}
\end{equation}
which holds independently of the gravitational field equations or the matter
content of the Universe. In the standard $\Lambda$CDM model, the jerk parameter
is constant and equal to unity, $j\equiv 1$, making deviations from this value
a natural kinematic indicator of deviations from the concordance scenario.

Motivated by the known asymptotic behaviors of the cosmic expansion, we
decompose the deceleration parameter as
\begin{equation}
q(z) = \frac{1}{2} + f(z) + g(z),
\end{equation}
Substituting this decomposition into
Eq.~\eqref{eq:jerk_identity_app} yields
\begin{equation}
j(z) = 1 + 3(f+g) + 2(f+g)^2 + (1+z)(f'+g'),
\end{equation}
where primes denote derivatives with respect to the redshift. This expression is
exact, but contains a non-linear term proportional to $f^2$, which prevents a
closed linear evolution equation for $f(z)$ without additional assumptions.

At late times, the early-time contribution becomes negligible, $g(z)\simeq 0$,
and the expression above reduces to
\begin{equation}
j(z)-1 \simeq 3f(z) + 2f^2(z) + (1+z)f'(z).
\end{equation}
In the strict limit for $\Lambda$CDM, where $j\equiv 1$, this equation admits the
solution $f(z)\propto (1+z)^{-3}$, corresponding to a monotonic decay and no
localized transition. Describing a smooth and localized late-time transition
therefore requires allowing for a controlled deviation from $j=1$.

From a kinematic point of view, we interpret $j(z)-1$ as an effective source term
and adopt a phenomenological closure in which the non-linear contribution
$2f^2$ is absorbed into this source. This leads to the approximate linear
relation
\begin{equation}
(1+z)f'(z) + 3f(z) = \Delta j(z),
\end{equation}
where $\Delta j(z)$ encodes the deviation from $\Lambda$CDM. A simple and stable
way to generate a localized transition is to prescribe the logarithmic
derivative of $f(z)$ as
\begin{equation}
\frac{d\ln f}{dz} = \frac{1}{1+z} - 2\beta(z+z_c),
\end{equation}
which integrates to
\begin{equation}
f(z) \propto (1+z)\exp\!\left[\beta\left(z_c^2-(z+z_c)^2\right)\right].
\end{equation}
This form corresponds to a Gaussian transition in the redshift modulated by the
factor $(1+z)$ and provides a smooth interpolation between a matter-dominated
regime and late-time acceleration. The parameter $\beta$ controls the width of
the transition, while $z_c$ sets its characteristic redshift.




\end{appendices}


\bibliography{References.bib}

\end{document}